\documentclass[amsmath, amssymb, aps, prb, twocolumn, notitlepage, longbibliography, superscriptaddress]{revtex4-2}

\usepackage{graphicx}
\usepackage{calc}
\usepackage{braket}
\usepackage{slashed}
\usepackage{wasysym}
\usepackage{dsfont}
\usepackage{float} 
\usepackage{amsthm,amsmath,amsfonts,amssymb,verbatim,color}
\usepackage{lipsum}
\usepackage{bm}
\usepackage{epsfig,slashed}
\usepackage{hyperref}
\usepackage{flafter}
\usepackage{booktabs}
\usepackage{makecell} 
\usepackage{bbm} 
\usepackage[export]{adjustbox} 
\usepackage{diagbox} 
\usepackage{makecell} 

\begin{document}	
	\title{Quasispins of vacancy defects in Ising chains with nearest- and next-to-nearest-neighbour
	 interactions}

	\author{Shijun Sun}
	\affiliation{Physics Department, University of California, Santa Cruz, California 95064, USA}

 \author{Arthur P. Ramirez}
	\affiliation{Physics Department, University of California, Santa Cruz, California 95064, USA}

 \author{Sergey Syzranov}
	\affiliation{Physics Department, University of California, Santa Cruz, California 95064, USA}

\begin{abstract}
Motivated by frustrated magnets and quasi-one-dimensional magnetic materials, we study the magnetic properties of 1D Ising chains with nearest-neighbour (NN) and weaker next-to-nearest neighbour (NNN) interactions in the presence of vacancy defects.
The effect of a vacancy on the magnetic susceptibility of 
a spin chain is two-fold: it reduces the length of the chain by an effective ``vacancy size'' and may also act as a free spin, a ``quasispin'', with a Curie-type 
$\chi_\text{quasi}=\langle S^2\rangle/T$ contribution to the susceptibility. 
In chains with antiferromagnetic short-range order,
the susceptibility of vacancy-free chains
is exponentially suppressed at low temperatures, and quasispins dominate the 
effect of impurities on the chains' magnetic properties.
For chains with antiferromagnetic NN interactions, the quasispin matches the value
$\langle S^2\rangle=1$ of the Ising spins in the chain
for ferromagnetic NNN interactions and vanishes for antiferromagnetic 
NNN interactions.
For chains with ferromagnetic short-range order, quasispin effects are insignificant
due to exponentially large low-temperature susceptibilities, 
and the dominant effect of a vacancy is effectively changing the length of the chain.
\end{abstract}

	
\maketitle

Impurities in magnetic materials are often seen as detrimental to magnetic order and collective quantum coherent phenomena such as quantum spin liquids.
Non-magnetic impurities can dilute the spins in a magnetic material, thus inhibiting long-range magnetic order. 
In materials with strong geometrical frustration~\cite{Ramirez:ReviewFrustrated},
small amounts of impurities lead to the formation of a spin glass state.

Despite their detrimental effects in some situations, impurities can also be used as a powerful probe of the low-temperature physics of pure materials. For example, an analysis of the spin-glass freezing temperature in geometrically frustrated magnets (GFMs) as a function of vacancy concentration suggests the existence of a ``hidden energy scale''~\cite{SyzranovRamirez}, at least an order of magnitude lower than the Weiss constant, at which random spin freezing occurs in very clean samples.


Vacancy defects can also have a significant effect on the magnetic susceptibility. 
In vacancy-free GFMs, the susceptibility above the hidden energy scale is described by the Curie-Weiss law $\chi(T)\propto (T+|\theta_{W}|)^{-1}$, where $\theta_{W}$ is the Weiss constant. 
Introducing vacancies can be naturally expected to re-scale the susceptibility by a factor of $\propto\left( N-N_\text{imp}\right)/N$ where $N_\text{imp}$ is the number of vacancies and $N$ is the 
total number of sites at which vacancies can be located.
At the same time,
introducing vacancies in a GFM leads to another
dramatic effect~\cite{SchifferDaruka,LaForge}: 
it gives rise to a Curie-like contribution $\propto 1/T$ to the susceptibility.
This suggests that, in addition
to diluting the bulk magnetic susceptibility in GFMs, vacancies lead to the emergence of degrees of freedom that behave as
free spins.

Such a Curie-like contribution has been attributed in Ref.~\cite{SchifferDaruka} to free ``orphan spins'',
i.e. spins 
disconnected from the other spins by quenched disorder (not necessarily vacancies).
The density of such orphan spins would, however, be too low~\cite{LaForge} in real materials to explain the experimentally observed susceptibility.
It has also been shown theoretically~\cite{Henley:HalfOrphans,SenDamleMoessner,PatilDamle:HalfOrphans} that ``half-orphan'' spins, i.e. spins located next to pairs of vacancies and thus not fully connected to the others,
have the susceptibility of free fractional spins.


The $\propto 1/T$ behaviour of the vacancy-induced magnetic susceptibility
can, however, be generically associated with individual vacancy defects. 
Indeed, for a system whose magnetisation 
$M_z$ along a certain direction commutes with the Hamiltonian,
the fluctuation-dissipation relation $\chi_{zz}(T)=\left< M_z^2\right>/T$ suggests 
the same temperature dependence as for the susceptibility of a free, possibly fractional
spin if $\left< M_z^2\right>\neq 0$ at $T=0$.

The existence of vacancy-associated magnetisation 
has also been theoretically demonstrated for defects in 2D Heisenberg antiferromagnets~\cite{SandvikDagottoScalapino,SachdevBuragonhainVojta,Sushkov,HoglundSandvik03,HoglundSandvik07,WollnyFritzVojta,WollnyVojta:vacancies}, classical vector spins~\cite{WollnyFritzVojta,WollnyVojta:vacancies,MaryasinZhitomirsky,Maryasin_2015}
and several-legged Heisenberg ladders~\cite{SandvikDagottoScalapino}. 
We note that the previous theoretical studies of vacancies in magnetic systems in dimensions $d>1$ focussed on magnetically ordered states.

For one-dimensional (1D) systems, analytical solutions are easily available, which 
often allows one to gain valuable insights into many-body phenomena. 
Vacancies in 1D chains with
nearest-neighbour (NN) interactions
also give rise to the $\propto 1/T$ contribution to the susceptibility~\cite{KatsuraTsujiyama,Wortis1974,Bogani,GoupalovMattis,ValkovShustin}.
In realistic magnetic systems, next-to-nearest-neighbour 
(NNN) interactions may significantly alter spin correlations around the vacancy and thus affect the susceptibility.
In addition to that, the presence of the NNN interactions, in general,
leads to frustration.

\begin{table}[htb!]
    \begin{center}
	\begin{tabular}{l| c| c}
			\toprule 
			\backslashbox[20mm]{NNN}{NN } & \makecell[c]{Antiferromagnetic \\ ($J_1>0$)}  & \makecell[c]{Ferromagnetic \\ ($J_1<0$)} \\
			\hline 
			\makecell[l]{Antiferro-\\magnetic\\ ($J_2>0$)}
			& \makecell{Quasispin $\langle S^{2}\rangle =0$;\\length increased by\\ $-b\approx e^{(2 J_{1}-6 J_{2})/T}$} & \makecell{Length reduced by \\ $ b\approx e^{\left(2|J_{1}|-4J_{2}\right)/T}$}\\
			\hline
			\makecell[l]{Ferro-\\magnetic\\ ($J_2<0$)}
            & \makecell{Quasispin $\langle S^{2}\rangle =1$;\\length reduced by\\ $b\approx e^{(2 J_{1}-2 J_{2})/T}$} & \makecell{Length reduced by\\ $b\approx e^{\left(2|J_{1}|-2J_{2}\right)/T}$}\\
			\bottomrule
		\end{tabular}
	\end{center}
	\caption{\label{Table} The leading effects of a single vacancy on 
		an Ising chain with the nearest-neighbour (NN) coupling $J_1$
		and next-to-nearest-neighbour (NNN) coupling $J_2$. 
		The vacancy effectively reduces the length of the chain
		by the ``size''	$b(T)$.
		In addition to that, a significant effect in chains with 
		antiferromagnetic NN interactions and ferromagnetic NNN interactions
		is the emergence of vacancy	``quasispins''.}
\end{table}

In this paper, we study the effect of vacancy defects 
on the magnetism of Ising chains with both NN and NNN interactions.
The presence of both NN and NNN interactions allows us to reveal some of the effects of vacancies on frustrated materials.
Our results can also be used to describe quasi-1D magnetic materials, such as
$\text{CoNb}_{2}\text{O}_{6}$\cite{Coldea:CoNb2O6,Kinross:CoNb2O6,sim:CoNb2O6} and $\text{CoCl}_{2}\cdot \text{D}_{2}\text{O}$\cite{Larsen:CoCl2} (with ferromagnetic ground states) and BCVO\cite{Zhao:BCVO,Niesen:BCVO,Zou:BCVO} (with an antiferromagnetic ground state). Although defects in these systems have not been studied yet, the results presented here would be applicable.
We compute analytically the magnetic susceptibility of such chains, revealing the contributions of both vacancies (``quasispins''), the bulk spins and their interplay.

{\it Summary of results.}
Due to a rapid decay of exchange interactions with distance, we focus on chains
with ratios of NN to NNN couplings $|J_1/J_2| > 2$. 
We find that the magnetic susceptibility of Ising chains with dilute vacancies is described by the formula	
	\begin{align}
		\chi(T)=\frac{\langle S^2\rangle }{T}N_\text{imp}
		+\frac{N-b(T) N_\text{imp}}{N}\chi_0(T),
		\label{Eq:QuasispinDefinition}
	\end{align}
where $\chi_0(T)$ is the susceptibility of a vacancy-free chain, and the length $b(T)$ is the effective ``vacancy size''.

As Eq.~\eqref{Eq:QuasispinDefinition} shows, the effect of vacancies is two-fold.
In general, a vacancy effectively reduces the length of the chain by the effective 
``vacancy size'' $b(T)$ and also acts as a ``quasispin'' of magnitude
$\sqrt{\langle S^2\rangle}$. Our results are summarised in 
Table~\ref{Table}. 
Such behaviour of magnetic susceptibility captures the physics 
of vacancy defects in GFMs.
The empirical dependence for the ``orphan''  spins proposed in Ref.~\cite{SchifferDaruka} 
corresponds to $b(T)=0$ and the susceptibility $\chi_0(T)$ described by the Curie-Weiss law.

In vacancy-free chains with antiferromagnetic NN interactions, the susceptibility 
$\chi_0(T)$
is exponentially suppressed and non-singular at $T=0$.
As a result, quasispin susceptibility,
the first term in Eq.~\eqref{Eq:QuasispinDefinition},
dominates the change $\partial \chi(T)/\partial N_\text{imp}$ of
the total susceptibility with vacancy concentration~\cite{Syzranov:quasispin,SyzranovRamirez}.
We find that for ferromagnetic NNN interactions, $\langle S^2\rangle=1$,
while for antiferromagnetic NNN interactions, $\langle S^2\rangle=0$.

In chains with ferromagnetic NN interactions, the susceptibility 
$\chi_0(T)$ is exponentially large at $T\rightarrow 0$ and more singular 
than the quasispin contribution. In that regime, the quasispins are insignificant. We compute the effective ``vacancy'' size $b(T)$ in all regimes.

{\it Qualitative interpretation.} The obtained
quasispin values can be understood as follows.
An infinite chain with a vacancy 
is equivalent to two half-infinite chains whose open ends are coupled by the NNN interaction at the vacancy. The end of a half-infinite chain may, in general, be expected to have its own magnetic moment $\mu$.
When the two ends are connected by ferromagnetic NNN interaction, their magnetic moments add up, resulting in the magnetic moment $2\mu$ of the vacancy. By contrast, antiferromagnetic NNN
interaction makes the magnetic moments of the open ends antialign, resulting in a zero magnetic moment of the vacancy, in accordance with the results of a direct calculation (see Table~\ref{Table}).

To complement this qualitative argument, in Appendix~\ref{Sec:finitechain},
we evaluate the magnetic moment of a free end of the spin chain explicitly
and obtain $\mu=\frac{1}{2}$, in accordance with vacancy quasispin 
$\left< S^2\right>=1$.


{\it Generic expressions for magnetic susceptibility.}
For the sufficiently fast decay of exchange interactions with distance we consider, the ground state is determined by the NN interactions:
all spins are aligned for $J_1<0$ and neighbouring spins 
are antialigned for $J_1>0$
regardless of the sign of the NNN coupling $J_2$.
	
We assume the temperature $T$ of the chain to be low and significantly exceeded by both the couplings $J_1$ and $J_2$,
as well as the energy
\begin{align}
	E_D=2|J_1|-4J_2>0
\end{align}
of a ``domain wall'',
i.e. the energy of a state obtained from the ground state by flipping all spins to the right 
of a given site.

According to the fluctuation-dissipation theorem, 
the susceptibility of chains with and without vacancies is given by
\begin{equation}
	{\chi}(T) = \frac{\langle M^2\rangle}{T}=\frac{1}{T}\sum_{i,j}\langle\sigma_{i}\sigma_{j}\rangle,
	\label{Eq:chi-correlation}
\end{equation}
where $M=\sum_i \sigma_i$ is the total magnetisation of the system; $\langle\ldots\rangle$ is averaging with respect to the thermal state of the system at temperature $T$; and the summation with respect to the indices $i$ and $j$ runs over all sites with spins. In Eq.~\ref{Eq:chi-correlation}, we used that the average magnetisation $\langle\sigma_i\rangle$ vanishes at each site.

\begin{figure}[ht!]
	\centering
	\includegraphics[width = 3.2in]{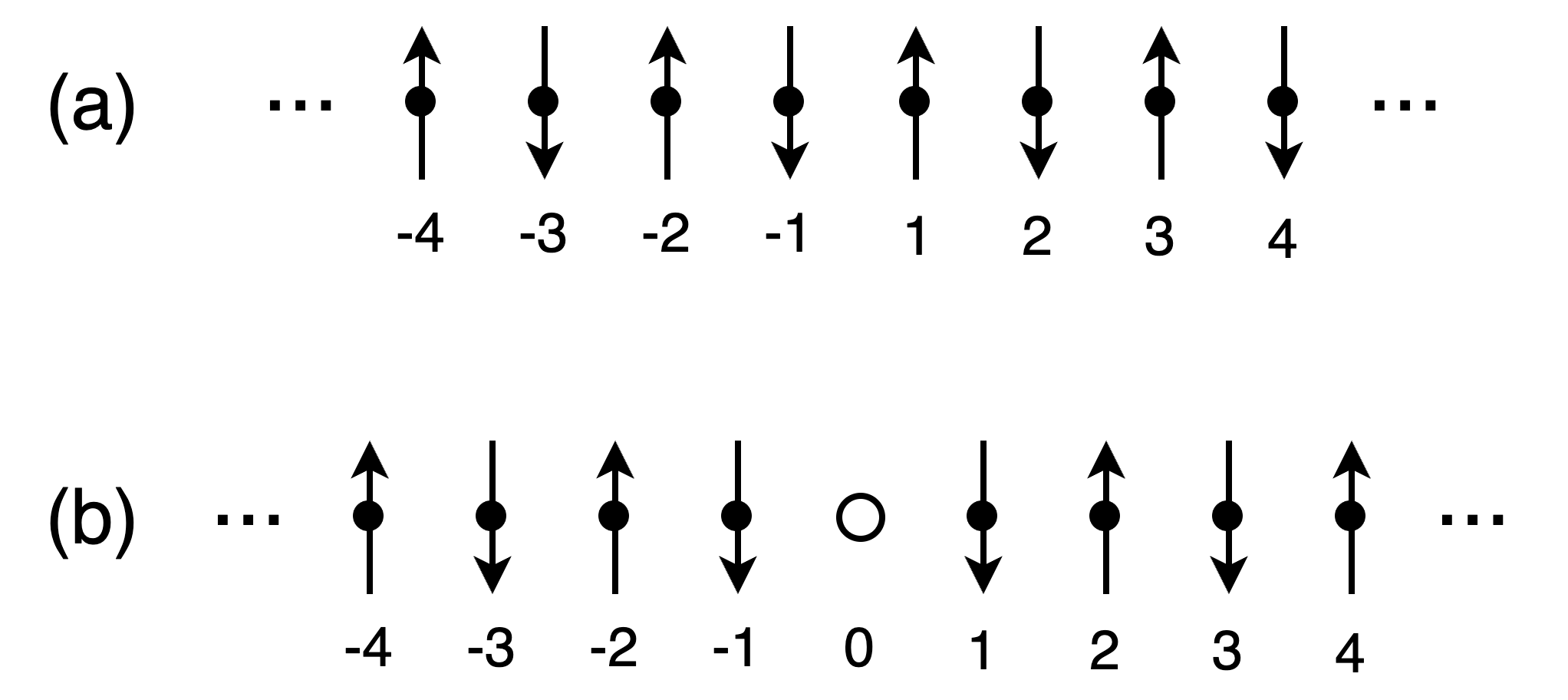}
	\caption{Spin configurations in the ground states of Ising chains with antiferromagnetic NN interactions and ferromagnetic NNN interactions.
		(a) A chain without a vacancy. (b) A chain with a vacancy.
		The labeling of the spins allows to conveniently compare the magnetic susceptibilities
		of the two chains and thus obtain the susceptibility of the vacancy (``quasispin'').}
	\label{fig:AFMNN+FMNNN}
\end{figure}

To compute the contribution of one vacancy to the susceptibility,
we utilise Eq.~\eqref{Eq:chi-correlation} to
obtain the susceptibilities $\chi(T)$
and $\chi_0(T)$ of the 
chains with and without a vacancy,
as shown in Fig.~\ref{fig:AFMNN+FMNNN}, and consider the 
difference 
\begin{align}
	\chi(T) - \chi_0(T)= & 2\beta\left(\langle\sigma_{-1}\sigma_{1}\rangle - \langle\sigma_{-1}\sigma_{1}\rangle_{0}\right)
	\nonumber \\
	&+4\beta \sum_{j=2}^{+\infty}\left(\langle\sigma_{1}\sigma_{j}\rangle-\langle\sigma_{1}\sigma_{j}\rangle_{0}\right) 
	\nonumber \\
	&+4\beta \sum_{j=2}^{+\infty}\left(\langle\sigma_{-1}\sigma_{j}\rangle-\langle\sigma_{-1}\sigma_{j}\rangle_{0}\right)
	\nonumber \\
	&+ 2\beta\sum_{i=-2}^{-\infty}\sum_{j=2}^{+\infty}\left(\langle\sigma_{i}\sigma_{j}\rangle -\langle\sigma_{i}\sigma_{j}\rangle_{0}\right),
	\label{Eq:differenceofsusceptibility}
\end{align}
where the averages
$\langle\cdots\rangle$ and $\langle\cdots\rangle_{0}$ are for the chains, respectively, with and without the vacancy, and
\begin{equation}
		\chi_0(T) \approx N T^{-1} e^{\mp E_D/T}
  \label{Eq:pseudospin0}
\end{equation}
is the susceptibility of the vacancy-free chain; ``$-$''
and ``$+$'' correspond, respectively, to ferromagnetic and
antiferromagnetic NN interactions.
In each of the sums in Eq.~\eqref{Eq:differenceofsusceptibility},
we extended one of the summation limits to infinity, assuming that the vacancy is located far (significantly further than the correlation length $\xi$) from the ends 
of the chain.

{\it Spin correlations in a gas of domain walls.}
The spin correlators in Eq.~\eqref{Eq:differenceofsusceptibility} can be conveniently evaluated by mapping the chain to a one-dimensional ideal gas of domain walls.
Indeed, the density of the domain walls (per site)
can be evaluated as $n_D\approx \exp\left({-E_D/T}\right)\ll 1$. 
This density significantly exceeds the density $n_{DD}\approx
n_D\exp\left(-2|J_1|/T\right)$ of double domain walls (domain walls on 
neighbouring sites). The approximation of an ideal gas of 
domain walls is, therefore, well justified in the considered limit of law
temperatures $T\ll |J_1|, |J_2|, E_D$.

The mapping to a gas of domain walls allows us to compute the correlators of magnetisation in Eq.~\eqref{Eq:differenceofsusceptibility}.
In the case of {\it antiferromagnetic NN interactions}, we obtain
(see Appendix~\ref{Sec:AFMNN+FMNNN} for details)
\begin{subequations}
    \begin{align}
        \langle\sigma_{-1}\sigma_{1} \rangle &= \frac{1-e^{2\beta J_{2}}}{1+e^{2\beta J_{2}}},
	\label{Corr1}\\
        \langle\sigma_{1}\sigma_{j} \rangle  &\approx  \frac{e^{-\beta E_{D}-2\beta J_{2}}-1}{1+e^{-\beta E_{D}-2\beta J_{2}}}\left(\frac{e^{-\beta E_{D}}-1}{1+e^{-\beta E_{D}}}\right)^{j-2},\\
        \langle\sigma_{-1}\sigma_{j} \rangle &\approx \langle\sigma_{-1}\sigma_{1}\rangle\langle\sigma_{1}\sigma_{j} \rangle ,\\
        \langle\sigma_{i}\sigma_{j} \rangle &\approx \langle\sigma_{i}\sigma_{-1}\rangle\langle\sigma_{-1}\sigma_{1}\rangle\langle\sigma_{1}\sigma_{j} \rangle,
		\label{Corr4}
	\end{align}
\end{subequations}
for $i\leq-2$ and $j\geq2$, where the spins in 
the chains with and without the vacancy are labelled as shown, respectively,
in Figs.~\ref{fig:AFMNN+FMNNN}a and \ref{fig:AFMNN+FMNNN}b.
In accordance with the chain's symmetry, the correlators
satisfy the relation
\begin{equation}
    \langle\sigma_{i}\sigma_{j}\rangle = \langle\sigma_{-i}\sigma_{-j}\rangle
\end{equation}
for all $i$ and $j$.

For antiferromagnetic NN and ferromagnetic NNN interactions ($J_{1}>0$, $J_{2}<0$), Eqs.~\eqref{Corr1}-\eqref{Corr4} 
and \eqref{Eq:chi-correlation} give the magnetic susceptibility 
\begin{align}
    \chi(T)- \chi_0(T) = T^{-1} &- T^{-1}e^{-2|J_{2}|/T} \nonumber\\
    &+o\left(T^{-1}e^{-2|J_{2}|/T}\right),
    \label{Eq:pseudospin1}
\end{align}
of the vacancy in the limit of low temperatures. 
In accordance with Eq.~\eqref{Eq:QuasispinDefinition},
the first term is the free-spin-type contribution corresponding to the 
quasispin $\langle S^{2}\rangle = 1$, and the second term 
describes an effective shortening of the chain by the 
``vacancy size'' $b(T)\approx e^{(2 J_{1}-2 J_{2})/T}$.

In the case of other signs of the couplings $J_1$ and $J_2$, 
the correlators $\langle \sigma_i \sigma_j\rangle$ can be computed similarly.
We present the details of the calculations of the susceptibility
of the chain with a vacancy 
for the cases $J_1>0, J_2>0$; $J_1<0; J_2>0$ and $J_1<0; J_2<0$
in, respectively, Appendices \ref{Sec:AFMNN+AFMNNN}, \ref{Sec:FMNN+AFMNNN}
and \ref{Sec:FMNN+FMNNN}. The results for the leading effects of the vacancy on the 
susceptibility are summarised in Table~\ref{Table}.

We emphasise that for ferromagnetic chains ($J_1<0$),
the susceptibility is exponentially large in temperature, $\chi(T)\propto \exp\left(E_D/T\right)$.
In this case, the main effect of the vacancy is changing the length of the chain by an effective vacancy
size $b(T)$ (see Table~\ref{Table} for the values of this length).

{\it Conclusion.} 
We have studied the effects of vacancy defects on the magnetic susceptibility of Ising chains with NN and (weaker) NNN interactions. 
We have shown, using analytical calculations, that the susceptibility of a chain with a vacancy is given, in general,
by a combination of two contributions:
the contribution of the bulk spins and the Curie-like contribution of vacancy ``quasispins''.

For chains with antiferromagnetic NN interactions, we found the quasispin of a vacancy to be, in the units of Ising spins, one (zero) for ferromagnetic (antiferromagnetic) NNN interactions. 
In the case of ferromagnetic NN interactions, the leading effect of a vacancy defect is the effective change of the chain length.

The susceptibilities of individual vacancies accurately 
describe the properties of magnetic materials in the limit of dilute vacancies, when the average distance between the defects significantly exceeds the correlation length $\xi$ of a clean material. 
At larger vacancy concentrations, quasispin-quasispin interactions may become essential.
The qualitative effect of such interactions may be particularly important in the case
of zero (or very small) quasispins of individual defects~\cite{Maryasin_2015}:
effective interactions may lead to a finite (large) collective magnetisation of the defects.

Another interesting aspect of quasispins that deserves 
investigation is their behaviour 
beyond linear response in magnetic field and behaviour 
in inhomogeneous magnetic fields. 
A cluster of spins in a GF system of a size smaller than the correlation length $\xi$
can be expected to behave similarly to a free spin whose magnetisation
is encoded by distinct ground states of the cluster. Unlike a conventional spin on a single site of the lattice, such a quasispin will be 
protected from local weak fluctuations of the field
due to the collective character of its magnetisation.

{\it Acknowledgements.} This work has been supported by
the NSF grant DMR-2218130.

	\bibliography{references}


	\onecolumngrid
	\vspace{2cm}
	
	\cleardoublepage

	
	\setcounter{equation}{0}
	\setcounter{figure}{0}
	\setcounter{enumiv}{0}

	\appendix

	\section{Chains with antiferromagnetic NN and ferromagnetic NNN interactions $J_{1}>0$, $J_{2}<0$}
	\label{Sec:AFMNN+FMNNN}

	\subsection{Mapping to a gas of domain walls}
	
In the ground state of a chain with antiferromagnetic NN and ferromagnetic NNN interactions, neighbouring spins are antiparallel.
Each excited state can be considered as a sequence of domain walls, i.e. pairs of neighbouring parallel spins, separating
antiferromagnetic domains.
Correlations between spins can be conveniently found by mapping the spin chain to a gas of domain walls.

In a system of size $r$, the partition function of the domain-wall excitations is given by
	\begin{equation}
		Z = 1 + C_{r}^{1}e^{-\beta E_{D}} +\left(C_{r}^{2}-C_{r-1}^{1}\right)e^{-2\beta E_{D}} + C_{r-1}^{1}e^{-2\beta E_{D}-4\beta J_{2}} +\cdots,
		\label{PartitionFunctionDW}
	\end{equation}
where 
\begin{align}
	E_D=2J_{1}-4J_{2}
\end{align}
is the energy of a single domain wall.
The second term in the right-hand side (rhs) of Eq.~\eqref{PartitionFunctionDW} is the contribution of a single domain wall that can be located anywhere.
The third term comes from two domain walls that are not located next to each other, with $C_{r}^{2}-C_{r-1}^{1}$ being the number of such configurations.
The fourth term comes from two domain walls next to each other, where $4J_2$ is the effective interaction energy of such walls.

In the limit of low temperatures, the domain-wall excitations are very sparse, and the system may be considered as an almost ideal gas of domain walls.
In this limit, configurations with domain walls located next to each other can be neglected.
Because the value of the product $\sigma_i\sigma_j$ of two spins distance $|i-j|=r$ apart is determined by the number of domain walls between them,
the correlator of those spins can be found as 
\begin{equation}
	\langle \sigma_{i}\sigma_{j}\rangle_{0} \approx(-1)^{r}\,\frac{1 - C_{r}^{1}e^{-\beta E_{D}} +C_{r}^{2}e^{-2\beta E_{D}}  -\cdots}{1 + C_{r}^{1}e^{-\beta E_{D}} +C_{r}^{2}e^{-2\beta E_{D}}  +\cdots} 
    = \left(\frac{e^{-\beta E_{D}}-1}{1+e^{-\beta E_{D}}}\right)^{r}.
	\label{Eq:SpincorrelationAFMNN+FMNNN}
\end{equation}
According to Eq.~\eqref{Eq:SpincorrelationAFMNN+FMNNN}, the correlation length in the spin chain is given by 
\begin{equation}
    \xi \approx \left[\log\left(\frac{1+e^{-\beta E_{D}}}{1-e^{-\beta E_{D}}}\right)\right]^{-1}\approx \frac{1}{2}e^{\beta E_{D}}.
\end{equation}
In this paper, we assume that the length of the chain $N$ significantly exceeds the correlation length $\xi$.

Equation~\eqref{Eq:SpincorrelationAFMNN+FMNNN} together with the fluctuation-dissipation relation~\eqref{Eq:chi-correlation}
give the magnetic susceptibility     
\begin{equation}
	\chi_{0}(T) = \beta \sum_{i,j}\langle\sigma_{i}\sigma_{j}\rangle_{0} \approx \beta N + 2\beta N \,\frac{e^{-\beta E_{D}}-1}{1+e^{-\beta E_{D}}}\,\left({1-\frac{e^{-\beta E_{D}}-1}{1+e^{-\beta E_{D}}}}\right)^{-1} = \beta N e^{-\beta E_{D}} = \beta N e^{-2\beta J_{1}+4\beta J_{2}}.
	\label{Eq:SusceptibilityAFMNN+FMNNN}
\end{equation}

	
\subsection{Magnetic susceptibility of a vacancy}

In this subsection, we compute the magnetic susceptibility of a chain with a vacancy.
The vacancy is assumed to be far away from the chain ends, with the distance much larger than the correlation length $\xi$. 
In what follows, we label the spins as shown in Fig.~\ref{fig:AFMNN+FMNNN}b.

For ferromagnetic NNN interactions considered here, the ground-state spin configuration in the chain with the vacancy is identical to that in the absence of the vacancy,
as shown in Figs.~\ref{fig:AFMNN+FMNNN}a and b. For excited states, the presence of the vacancy modifies the domain-wall energies near and at the vacancy.
A domain wall at the location of the vacancy, i.e. an excitation corresponding to anti-aligned spins at sites 
$-1$ and $1$, has an energy of $2|J_{2}|$.
A domain-wall excitation between sites $1$ and $2$ or between sites $-1$ and $-2$ has an energy of $2J_{1}-2J_{2}$. 
The energy of domain walls at the other locations is unchanged by the vacancy, $E_{D} = 2J_{1}-4J_{2}$.

Using these excitation energies, we obtain the correlators of spins in the chain with a vacancy:
 \begin{subequations}
	\begin{align}
		\langle\sigma_{-1}\sigma_{1} \rangle &= \frac{1-e^{2\beta J_{2}}}{1+e^{2\beta J_{2}}}, \label{Eq:correlationvacancyAFMFM1}\\
        \langle\sigma_{1}\sigma_{j} \rangle  &\approx  \frac{e^{-2\beta J_{1}+2\beta J_{2}}-1}{1+e^{-2\beta J_{1}+2\beta J_{2}}}\left(\frac{e^{-\beta E_{D}}-1}{1+e^{-\beta E_{D}}}\right)^{j-2},\\
        \langle\sigma_{-1}\sigma_{j} \rangle &\approx \frac{1-e^{2\beta J_{2}}}{1+e^{2\beta J_{2}}}\,\frac{e^{-2\beta J_{1}+2\beta J_{2}}-1}{1+e^{-2\beta J_{1}+2\beta J_{2}}} \left(\frac{e^{-\beta E_{D}}-1}{1+e^{-\beta E_{D}}}\right)^{j-2},\\
        \langle\sigma_{i}\sigma_{j} \rangle &\approx\frac{1-e^{2\beta J_{2}}}{1+e^{2\beta J_{2}}}\left(\frac{e^{-2\beta J_{1}+2\beta J_{2}}-1}{1+e^{-2\beta J_{1}+2\beta J_{2}}}\right)^{2}\left(\frac{e^{-\beta E_{D}}-1}{1+e^{-\beta E_{D}}}\right)^{j-i-4}, \label{Eq:correlationvacancyAFMFM4}
	\end{align}
\end{subequations}
for $i\leq -2$, $j\geq2$. 
For $i,j\leq -2$ and $i,j\geq2$, the correlators are unaltered by the presence of the vacancy and are given by Eq.~\eqref{Eq:SpincorrelationAFMNN+FMNNN}.
All the correlators satisfy the symmetry relation $\langle\sigma_{i}\sigma_{j}\rangle = \langle\sigma_{-i}\sigma_{-j}\rangle$.
When obtaining the correlators~\eqref{Eq:correlationvacancyAFMFM1}-\eqref{Eq:correlationvacancyAFMFM4}, we used the approximation of low temperature and, similar to the case of a vacancy-free chain,
neglected the probabilities of having two domain walls next to each other.

	The magnetic susceptibility of the chain is given by
	\begin{align}
		\chi(T) &= \beta \sum_{i,j\neq 0}\langle\sigma_{i}\sigma_{j}\rangle \nonumber\\
		&= \beta\sum_{i\neq 0}\langle\sigma_{i}^{2}\rangle+2\beta \left(\langle\sigma_{-1}\sigma_{1}\rangle + 2\sum_{j=2}^{\infty}\langle\sigma_{1}\sigma_{j}\rangle+2\sum_{j=2}^{\infty}\langle\sigma_{-1}\sigma_{j}\rangle + \sum_{i=-2}^{-\infty}\sum_{j=2}^{\infty}\langle\sigma_{i}\sigma_{j}\rangle + 2\sum_{i,j=2,\,i\neq j}^{\infty}\langle\sigma_{i}\sigma_{j}\rangle\right).
		\label{Eq:SusceptibilitywithVacancy}
	\end{align}
	
The magnetic susceptibility of the vacancy is given by the difference in the susceptibilities of the chain with and without the vacancy:
	\begin{align}
		\chi(T) - \chi_{0}(T)&= 2\beta\left(\langle\sigma_{-1}\sigma_{1}\rangle - \langle\sigma_{-1}\sigma_{1}\rangle_{0}\right)+4\beta \sum_{j=2}^{\infty}\left(\langle\sigma_{1}\sigma_{j}\rangle-\langle\sigma_{1}\sigma_{j}\rangle_{0}\right) \nonumber\\
		&\quad+4\beta \sum_{j=2}^{\infty}\left(\langle\sigma_{-1}\sigma_{j}\rangle-\langle\sigma_{-1}\sigma_{j}\rangle_{0}\right)+ 2\beta\sum_{i=-2}^{-\infty}\sum_{j=2}^{\infty}\left(\langle\sigma_{i}\sigma_{j}\rangle -\langle\sigma_{i}\sigma_{j}\rangle_{0}\right),
		\label{ChiDifferenceAppA}
	\end{align}
where the correlators $\langle
\sigma_i\sigma_j\rangle_0$ in a vacancy-free system are given by Eq.~\eqref{Eq:SpincorrelationAFMNN+FMNNN} and, in terms of the labels 
chosen in this subsection [cf. Fig.~\ref{fig:AFMNN+FMNNN}] can be rewritten as
    \begin{subequations}
        \begin{align}
            \langle\sigma_{i}\sigma_{j} \rangle_{0} &\approx \left(\frac{e^{-\beta E_{D}}-1}{1+e^{-\beta E_{D}}}\right)^{j-i-1}, \quad \text{for} \,\, i\leq -1, \,\, j\geq1;
		\label{Eq:correlationnovacancyAFM1}\\
            \langle\sigma_{i}\sigma_{j} \rangle_{0} &\approx \left(\frac{e^{-\beta E_{D}}-1}{1+e^{-\beta E_{D}}}\right)^{|j-i|}, \quad \text{for}\,\, i,j\leq-1 \,\,\text{and}\,\, i,j\geq1.
		\label{Eq:correlationnovacancyAFM2}
        \end{align}
    \end{subequations}

Utilising Eqs.~\eqref{Eq:correlationvacancyAFMFM1}-\eqref{Eq:correlationvacancyAFMFM4}, \eqref{Eq:correlationnovacancyAFM1}-\eqref{Eq:correlationnovacancyAFM2}
and \eqref{ChiDifferenceAppA}, we obtain the magnetic susceptibility of the vacancy:
\begin{align}
	\chi(T) - \chi_{0}(T) &\approx 2\beta \left(\frac{1-e^{2\beta J_{2}}}{1+e^{2\beta J_{2}}}- \frac{e^{-\beta E_{D}}-1}{1+e^{-\beta E_{D}}}\right) \nonumber\\
	&\quad + 4\beta \left(\frac{1}{1+e^{2\beta J_{2}}}\frac{e^{-2\beta J_{1}+2\beta J_{2}}-1}{1+e^{-2\beta J_{1}+2\beta J_{2}}}-\frac{e^{-\beta E_{D}}}{1+e^{-\beta E_{D}}}\frac{e^{-\beta E_{D}}-1}{1+e^{-\beta E_{D}}}\right)\left(1+e^{-\beta E_{D}}\right)\nonumber\\
	&\quad + \frac{\beta}{2}\left[\frac{1-e^{2\beta J_{2}}}{1+e^{2\beta J_{2}}}\left(\frac{e^{-2\beta J_{1}+2\beta J_{2}}-1}{1+e^{-2\beta J_{1}+2\beta J_{2}}}\right)^{2}-\left(\frac{e^{-\beta E_{D}}-1}{1+e^{-\beta E_{D}}}\right)^{3}\right]\left(1+e^{-\beta E_{D}}\right)^{2}.
\end{align}
	At low temperatures, $T \ll 2J_{1},\,2|J_2|$, 
	\begin{equation}
		\chi(T)-\chi_{0}(T) = \beta -\beta e^{-2\beta |J_{2}|}+6\beta e^{-2\beta J_{1}-2\beta |J_{2}|} + o\left(\beta e^{-2\beta J_{1} - 2\beta |J_{2}|}\right)
		\label{Eq:pseudospin}
	\end{equation}
    The leading order contributions give equation \eqref{Eq:pseudospin1} in the main text and corresponds to the vacancy quasispin $\langle S^{2}\rangle =1$ and the vacancy size $b(T) \approx e^{2\beta J_{1}-2\beta J_{2}}$.


\section{Chains with antiferromagnetic NN and NNN interactions, $J_{1}>0$, $J_{2}>0$}
	\label{Sec:AFMNN+AFMNNN}

\begin{figure}[ht!]
 \centering
 \includegraphics[width = 3in]{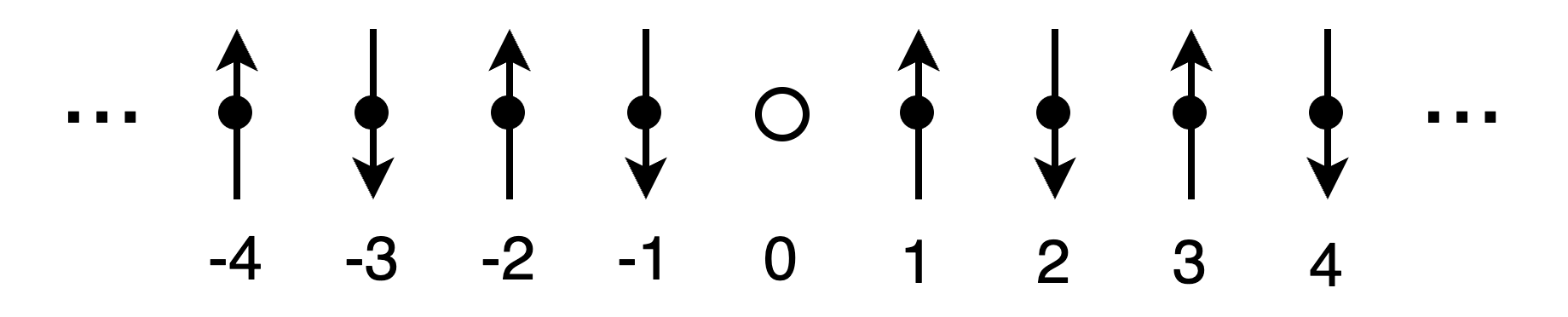}
 \caption{The ground state of chain with a vacancy antiferromagnetic 
 	NN and NNN interactions. Half of the spins are flipped relative to the chain's vacancy-free counterpart shown in Fig.~\ref{fig:AFMNN+FMNNN}a.
  }
 \label{fig:FlippedAFchain}
\end{figure}
 
This section deals with the case of a chain with antiferromagnetic NN and antiferromagnetic NNN interactions ($J_{1}<0$, $J_{2}>0$).
Due to strong antiferromagnetic NN interactions, the ground state of such a chain is still antiferromagnetic, with neighbouring spins antiparallel.
Low-temperature spin correlators in the absence of a vacancy are given by
\begin{equation}
    \langle \sigma_{i}\sigma_{j}\rangle_{0}\approx\left(\frac{e^{-\beta E_{D}}-1}{1+e^{-\beta E_{D}}}\right)^{|j-i|},
    \label{Eq:SpincorrelationAFMNN+AFMNNN}
\end{equation}
and the energy of a domain wall given by
    \begin{equation}
        E_{D} = 2J_{1}-4J_{2}.
    \end{equation}
	
The magnetic susceptibility of a vacancy-free chain is given by
	\begin{equation}
		\chi_{0}(T) = \beta \sum_{i,j}\langle\sigma_{i}\sigma_{j}\rangle_{0} \approx \beta N e^{-2\beta J_{1}+4\beta J_{2}}.
		\label{Eq:SusceptibilityAFMNN+AFMNNN}
	\end{equation}

Similar to the case considered in Appendix~\ref{Sec:AFMNN+FMNNN},
to find the susceptibility of a vacancy, we consider 
a chain with a vacancy located sufficiently far away from the ends. The spins are labeled as shown in Fig.~\ref{fig:FlippedAFchain}.

The ground state has half of the spins flipped on one side of the vacancy
relative to the other side, as shown in Figs.~\ref{fig:AFMNN+FMNNN}a and \ref{fig:FlippedAFchain}. A domain wall excitation at the location of the vacancy, i.e. an excitation corresponding to parallel spins at sites $-1$ and $1$, has an energy of $2J_{2}$. A domain wall excitation between sites $1$ and $2$ or between sites $-1$ and $-2$ has an energy of $2J_{1}-2J_{2}$. The energy of domain walls at other locations is unaltered by the vacancy, $E_{D} = 2J_{1}-4J_{2}$.
	
Spins correlators in the chain with a vacancy are given by
    \begin{subequations}
        \begin{align}
            \langle\sigma_{-1}\sigma_{1} \rangle &= \frac{e^{-2\beta J_{2}}-1}{1+e^{-2\beta J_{2}}},\label{Eq:correlationvacancyAFMAFM1}\\
            \langle\sigma_{1}\sigma_{j} \rangle &\approx  \frac{e^{-2\beta J_{1}+2\beta J_{2}}-1}{1+e^{-2\beta J_{1}+2\beta J_{2}}}\left(\frac{e^{-\beta E_{D}}-1}{1+e^{-\beta E_{D}}}\right)^{j-2}, \\
            \langle\sigma_{-1}\sigma_{j} \rangle &\approx \frac{e^{-2\beta J_{2}}-1}{1+e^{-2\beta J_{2}}}\,\frac{e^{-2\beta J_{1}+2\beta J_{2}}-1}{1+e^{-2\beta J_{1}+2\beta J_{2}}} \left(\frac{e^{-\beta E_{D}}-1}{1+e^{-\beta E_{D}}}\right)^{j-2},\\
            \langle\sigma_{i}\sigma_{j} \rangle &\approx \frac{e^{-2\beta J_{2}}-1}{1+e^{-2\beta J_{2}}}\left(\frac{e^{-2\beta J_{1}+2\beta J_{2}}-1}{1+e^{-2\beta J_{1}+2\beta J_{2}}}\right)^{2}\left(\frac{e^{-\beta E_{D}}-1}{1+e^{-\beta E_{D}}}\right)^{j-i-4}, \label{Eq:correlationvacancyAFMAFM4}
        \end{align}
    \end{subequations}
	for $i\leq -2$, $j\geq2$. For $i,j\leq -2$ and $i,j\geq2$, $\langle\sigma_{i}\sigma_{j} \rangle$, the correlators  are unchanged by the presence of the vacancy and are given by \eqref{Eq:SpincorrelationAFMNN+AFMNNN}. All the correlators satisfy symmetry relation $\langle\sigma_{i}\sigma_{j}\rangle = \langle\sigma_{-i}\sigma_{-j}\rangle$. When obtaining the correlators \eqref{Eq:correlationvacancyAFMAFM1}-\eqref{Eq:correlationvacancyAFMAFM4}, we have again used the approximation of low temperature and neglected the probabilities of having two domain walls next to each other. These results are the same as in the case of ferromagnetic NNN interactions [cf. Eqs.~\eqref{Eq:correlationvacancyAFMFM1}-\eqref{Eq:correlationvacancyAFMFM4}].
	
	The magnetic susceptibility of the vacancy is given by 
	\begin{align}
		\chi(T) - \chi_{0}(T)&= 2\beta\left(\langle\sigma_{-1}\sigma_{1}\rangle - \langle\sigma_{-1}\sigma_{1}\rangle_{0}\right)+4\beta \sum_{j=2}^{\infty}\left(\langle\sigma_{1}\sigma_{j}\rangle-\langle\sigma_{1}\sigma_{j}\rangle_{0}\right) 
		\nonumber\\
		&\quad+4\beta \sum_{j=2}^{\infty}\left(\langle\sigma_{-1}\sigma_{j}\rangle-\langle\sigma_{-1}\sigma_{j}\rangle_{0}\right)+ 2\beta\sum_{i=-2}^{-\infty}\sum_{j=2}^{\infty}\left(\langle\sigma_{i}\sigma_{j}\rangle -\langle\sigma_{i}\sigma_{j}\rangle_{0}\right),
		\nonumber \\
		&\approx 2\beta \left(\frac{e^{-2\beta J_{2}}-1}{1+e^{-2\beta J_{2}}}- \frac{e^{-\beta E_{D}}-1}{1+e^{-\beta E_{D}}}\right) \nonumber\\
		&\quad + 4\beta \left(\frac{e^{-2\beta J_{2}}}{1+e^{-2\beta J_{2}}}\,\frac{e^{-2\beta J_{1}+2\beta J_{2}}-1}{1+e^{-2\beta J_{1}+2\beta J_{2}}} - \frac{e^{-\beta E_{D}}}{1+e^{-\beta E_{D}}}\frac{e^{-\beta E_{D}}-1}{1+e^{-\beta E_{D}}}\right)\left(1+e^{-\beta E_{D}}\right)\nonumber\\
		&\quad + \frac{\beta}{2}\left[\frac{e^{-2\beta J_{2}}-1}{1+e^{-2\beta J_{2}}}\left(\frac{e^{-2\beta J_{1}+2\beta J_{2}}-1}{1+e^{-2\beta J_{1}+2\beta J_{2}}}\right)^{2}-\left(\frac{e^{-\beta E_{D}}-1}{1+e^{-\beta E_{D}}}\right)^{3}\right]\left(1+e^{-\beta E_{D}}\right)^{2}.
	\end{align}
	At low temperatures, $T \ll 2J_2, \,2J_1-4J_2$,
	\begin{equation}
		\chi(T) - \chi_{0}(T) = \beta e^{-2\beta J_{2}} + o\left(\beta e^{-2\beta J_{2}}\right)
		\label{Eq:diffSusceptibilityAFM},
	\end{equation}
	The exponential suppression of the $\chi(T)-\chi_0(T)$
	corresponds to a vanishing pseudospin of the vacancy,
	$\langle S^{2}\rangle =0$. The leading term corresponds to an increase in the chain length by $-b(T) \approx e^{2\beta J_{1} - 6\beta J_{2}}$.

\section{Chains with ferromagnetic NN and antiferromagnetic NNN interactions, $J_{1}<0$, $J_{2}>0$}
\label{Sec:FMNN+AFMNNN}

In this section,
we provide the details of the calculation of the magnetic susceptibility of a chain with ferromagnetic NN interactions and antiferromagnetic NNN interactions ($J_{1}<0$, $J_{2}>0$), following similar analysis in
in Appendix.~\ref{Sec:AFMNN+FMNNN} for a chain with $J_1>0$ and $J_2<0$.
 
In the ground state, all spins are parallel, owing to large NN interactions, $|J_1|>2J_2$.
In excited states, the system splits into ferromagnetic domains, with
domain walls corresponding to pairs of antiparallel neighbouring spins.
Low-temperature correlators of spins can be obtained similarly to those
in Appendix.~\ref{Sec:AFMNN+FMNNN} and are given by
	\begin{equation}
		\langle \sigma_{i}\sigma_{j}\rangle_{0}\approx \frac{1 - C_{r}^{1}e^{-\beta E_{D}} +C_{r}^{2}e^{-2\beta E_{D}}  -\cdots}{1 + C_{r}^{1}e^{-\beta E_{D}} +C_{r}^{2}e^{-2\beta E_{D}}  +\cdots}= \left(\frac{1-e^{-\beta E_{D}}}{1+e^{-\beta E_{D}}}\right)^{|i-j|}.
		\label{Eq:SpincorrelationFMNN+AFMNNN}
	\end{equation}
    where
    \begin{equation}
        E_{D} = -2J_{1}-4J_{2}
    \end{equation}
    is the energy of a domain wall.

    Equation \eqref{Eq:SpincorrelationFMNN+AFMNNN} together with the fluctuation-dissipation relation \eqref{Eq:chi-correlation} gives the magnetic susceptibility
	\begin{equation}
		\chi_{0}(T) = \beta \sum_{i,j}\langle\sigma_{i}\sigma_{j}\rangle_{0} \approx \beta N + 2\beta N \,\frac{1-e^{-\beta E_{D}}}{1+e^{-\beta E_{D}}}\,\left({1-\frac{1-e^{-\beta E_{D}}}{1+e^{-\beta E_{D}}}}\right)^{-1} = \beta N e^{\beta E_{D}} = \beta N e^{-2\beta J_{1}-4\beta J_{2}}.
		\label{Eq:SusceptibilityFMNN+AFMNNN}
	\end{equation}
    The magnetic susceptibility for a ferromagnetic chain is exponentially large.
	
	\begin{figure}[ht]
		\centering
		\includegraphics[width = 3in]{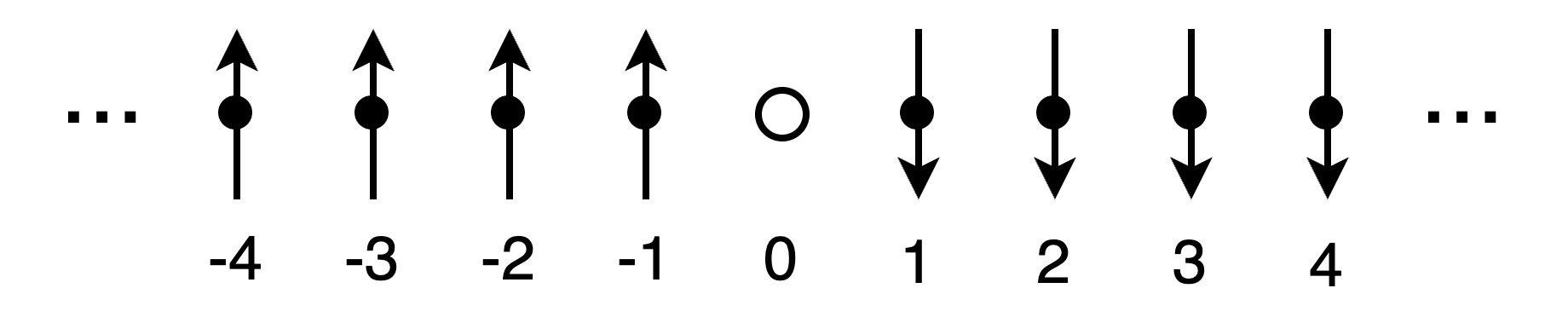}
		\caption{The ground state of chain with strong ferromagnetic NN and antiferromagnetic NNN interactions with a vacancy.}
		\label{fig:FMNN+AFMNNNwithvacancy}
	\end{figure}

\begin{figure}[ht]
		\centering
		\includegraphics[width = 2.7in]{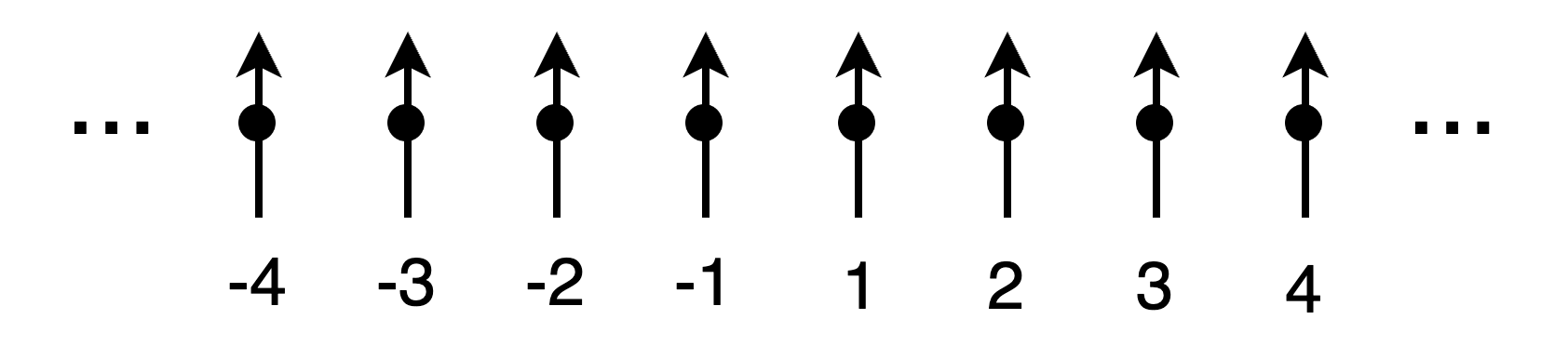}
		\caption{Ising chain with ferromagnetic NN interactions and antiferromagnetic NNN interactions. The chain has no vacancy and has the same number of spins as the chain with a vacancy. We compare the magnetic susceptibility of this chain to the chain with a vacancy Fig.~\ref{fig:FMNN+AFMNNNwithvacancy}.}
		\label{fig:FMNN+AFMNNNnovacancy}
	\end{figure}

For a chain with antiferromagnetic NNN interactions considered here, the ground-state spin configuration in a chain with a vacancy 
has half of the spins on one side of the vacancy antiparallel to
the spins on the other half, as shown in Figs.~\ref{fig:FMNN+AFMNNNwithvacancy} and \ref{fig:FMNN+AFMNNNnovacancy}. For excited states, the presence of the vacancy modifies the domain-wall energies near and at the vacancy. A domain wall excitation at the location of the vacancy, i.e. an excitation corresponding to aligned spins at sites $-1$ and $1$, has an energy of $2J_{2}$. A domain-wall excitation between sites $1$ and $2$ or between sites $-1$ and $-2$ has an energy of $-2J_{1}-2J_{2}$. The energy of domain walls at other locations is unchanged by the vacancy, $E_{D} = -2J_{1}-4J_{2}$.
	
These excitation energies result in the following spin correlators:
\begin{subequations}
    \begin{align}
       \langle\sigma_{-1}\sigma_{1} \rangle &= \frac{e^{-2\beta J_{2}}-1}{1+e^{-2\beta J_{2}}},\label{Eq:correlationvacancyFMAFM1}\\
       \langle\sigma_{1}\sigma_{j} \rangle  &\approx  \frac{1-e^{2\beta J_{1}+2\beta J_{2}}}{1+e^{2\beta J_{1}+2\beta J_{2}}} \left(\frac{1-e^{-\beta E_{D}}}{1+e^{-\beta E_{D}}}\right)^{j-2},\\
       \langle\sigma_{-1}\sigma_{j} \rangle &\approx\frac{e^{-2\beta J_{2}}-1}{1+e^{-2\beta J_{2}}}\,\frac{1-e^{2\beta J_{1}+2\beta J_{2}}}{1+e^{2\beta J_{1}+2\beta J_{2}}} \left(\frac{1-e^{-\beta E_{D}}}{1+e^{-\beta E_{D}}}\right)^{j-2},\\
       \langle\sigma_{i}\sigma_{j} \rangle &\approx\, \frac{e^{-2\beta J_{2}}-1}{1+e^{-2\beta J_{2}}}\left(\frac{1-e^{2\beta J_{1}+2\beta J_{2}}}{1+e^{2\beta J_{1}+2\beta J_{2}}}\right)^{2}\left(\frac{1-e^{-\beta E_{D}}}{1+e^{-\beta E_{D}}}\right)^{j-i-4},\label{Eq:correlationvacancyFMAFM4}
   \end{align}
\end{subequations}
	for $i\leq -2$, $j\geq2$. For $i,j\leq -2$ and $i,j\geq2$, $\langle\sigma_{i}\sigma_{j} \rangle$, the correlators  are unchanged by the presence of the vacancy and are given by \eqref{Eq:SpincorrelationFMNN+AFMNNN}. All the correlators satisfy symmetry relation $\langle\sigma_{i}\sigma_{j}\rangle = \langle\sigma_{-i}\sigma_{-j}\rangle$. 
	
	The magnetic susceptibility of the vacancy is given by 
	\begin{align}
		\chi(T) - \chi_{0}(T)&= 2\beta\left(\langle\sigma_{-1}\sigma_{1}\rangle - \langle\sigma_{-1}\sigma_{1}\rangle_{0}\right)+4\beta \sum_{j=2}^{\infty}\left(\langle\sigma_{1}\sigma_{j}\rangle-\langle\sigma_{1}\sigma_{j}\rangle_{0}\right) \nonumber\\
		&\quad+4\beta \sum_{j=2}^{\infty}\left(\langle\sigma_{-1}\sigma_{j}\rangle-\langle\sigma_{-1}\sigma_{j}\rangle_{0}\right)+ 2\beta\sum_{i=-2}^{-\infty}\sum_{j=2}^{\infty}\left(\langle\sigma_{i}\sigma_{j}\rangle -\langle\sigma_{i}\sigma_{j}\rangle_{0}\right).
        \label{ChiDifferenceAppC}
	\end{align}
    where the correlators $\langle\sigma_i\sigma_j\rangle_0$ in a vacancy-free system are given by Eq.~\eqref{Eq:SpincorrelationFMNN+AFMNNN} and, in terms of the labels chosen in this subsection [cf. Fig.~\ref{fig:FMNN+AFMNNNnovacancy}] can be rewritten as
    \begin{subequations}
        \begin{align}
            \langle\sigma_{i}\sigma_{j} \rangle_{0} &\approx \left(\frac{1-e^{-\beta E_{D}}}{1+e^{-\beta E_{D}}}\right)^{j-i-1}, \quad \text{for} \,\, i\leq -1, \,\, j\geq1;
		\label{Eq:correlationnovacancyFM1}\\
            \langle\sigma_{i}\sigma_{j} \rangle_{0} &\approx \left(\frac{1-e^{-\beta E_{D}}}{1+e^{-\beta E_{D}}}\right)^{|j-i|}, \quad \text{for}\,\, i,j\leq-1 \,\,\text{and}\,\, i,j\geq1.
		\label{Eq:correlationnovacancyFM2}
        \end{align}
    \end{subequations}
    Utilising Eqs.~\eqref{Eq:correlationvacancyFMAFM1}-\eqref{Eq:correlationvacancyFMAFM4}, \eqref{Eq:correlationnovacancyFM1}-\eqref{Eq:correlationnovacancyFM2} and \eqref{ChiDifferenceAppC}, we obtain the magnetic susceptibility of the vacancy:
	\begin{align}
		\chi(T) - \chi_{0}(T) &\approx 2\beta \left(\frac{e^{-2\beta J_{2}}-1}{1+e^{-2\beta J_{2}}} - \frac{1-e^{-\beta E_{D}}}{1+e^{-\beta E_{D}}}\right)\nonumber\\
		&\quad + 4\beta\left[\frac{e^{-2\beta J_{2}}}{1+e^{-2\beta J_{2}}}\,\frac{1-e^{2\beta J_{1}+2\beta J_{2}}}{1+e^{2\beta J_{1}+2\beta J_{2}}} - \frac{1}{1+e^{-\beta E_{D}}}\,\frac{1-e^{-\beta E_{D}}}{1+e^{-\beta E_{D}}}\right]\left(1+e^{\beta E_{D}}\right)\nonumber\\
		&\quad+\frac{\beta}{2}\left[\frac{e^{-2\beta J_{2}}-1}{1+e^{-2\beta J_{2}}}\left(\frac{1-e^{2\beta J_{1}+2\beta J_{2}}}{1+e^{2\beta J_{1}+2\beta J_{2}}}\right)^{2}-\left(\frac{1-e^{-\beta E_{D}}}{1+e^{-\beta E_{D}}}\right)^{3}\right]\left(1+e^{\beta E_{D}}\right)^{2}
	\end{align}
	At low temperatures, $T \ll 2J_2, \,2|J_1|-4J_2$
	\begin{equation}
		\chi(T) - \chi_{0}(T) = -\beta e^{-4\beta J_{1} - 8\beta J_{2}} +\beta e^{-4\beta J_{1}-10\beta J_{2}} + o\left(\beta e^{-4\beta J_{1}-10\beta J_{2}}\right)
		\label{Eq:diffSusceptibilityFM}
	\end{equation}
    The presence of a spin vacancy has an exponentially large effect on magnetic susceptibility. The leading order effect is equivalent to reducing the length of the chain by twice the correlation length $b(T)\approx 2\xi \approx e^{-2\beta J_{1}-4\beta J_{2}}$.

	\section{Chains with ferromagnetic NN and NNN interactions, $J_{1}<0$, $J_{2}<0$}
	\label{Sec:FMNN+FMNNN}
	
	In this section, we consider the case of an Ising chain with ferromagnetic NN and NNN interactions, $J_{1}<0$, $J_{2}<0$.
	
	In the ground state, all spins are parallel to each other,
	while excited states correspond to sequences of ferromagnetic domain walls.
	Low-temperature spin correlators are given by
	\begin{equation}
		\langle \sigma_{i}\sigma_{j}\rangle_{0} \approx \left(\frac{1-e^{-\beta E_{D}}}{1+e^{-\beta E_{D}}}\right)^{|j-i|}.
		\label{Eq:SpincorrelationFMNN+FMNNN}
	\end{equation}
    where 
    \begin{equation}
        E_{D} = -2J_{1}-4J_{2}
    \end{equation}
    is the energy of the domain wall.  

    The magnetic susceptibility of a vacancy-free chain is given by
	\begin{equation}
		\chi_{0}(T) = \beta \sum_{i,j}\langle\sigma_{i}\sigma_{j}\rangle_{0} \approx \beta N e^{-2\beta J_{1}-4\beta J_{2}}
		\label{Eq:SusceptibilityFMNN+FMNNN}
	\end{equation}
	and is exponentially large at low temperatures.
	
In the presence of a vacancy, all spins in the chain are still parallel. 
%
%
A domain wall at the location of the vacancy, i.e. an excitation corresponding to antiparallel spins at sites $-1$ and $1$, has an energy of $2|J_{2}|$. A domain-wall excitation between sites $1$ and $2$ or between sites $-1$ and $-2$ has an energy of $-2J_{1}-2J_{2}$.  The energy of domain walls at the other locations is unchanged by the vacancy, $E_{D} = -2J_{1}-4J_{2}$.
Low-temperature correlators of the spins in the chain are given by
    \begin{subequations}
        \begin{align}
            \langle\sigma_{-1}\sigma_{1} \rangle &= \frac{1-e^{2\beta J_{2}}}{1+e^{2\beta J_{2}}}, \label{Eq:correlationvacancyFMFM1}\\
            \langle\sigma_{1}\sigma_{j} \rangle  &\approx \frac{1-e^{2\beta J_{1}+2\beta J_{2}}}{1+e^{2\beta J_{1}+2\beta J_{2}}}\left(\frac{1-e^{-\beta E_{D}}}{1+e^{-\beta E_{D}}}\right)^{j-2},\\
            \langle\sigma_{-1}\sigma_{j} \rangle &\approx \frac{1-e^{2\beta J_{2}}}{1+e^{2\beta J_{2}}}\,\frac{1-e^{2\beta J_{1}+2\beta J_{2}}}{1+e^{2\beta J_{1}+2\beta J_{2}}} \left(\frac{1-e^{-\beta E_{D}}}{1+e^{-\beta E_{D}}}\right)^{j-2},\\
            \langle\sigma_{i}\sigma_{j} \rangle &\approx\frac{1-e^{2\beta J_{2}}}{1+e^{2\beta J_{2}}}\left(\frac{1-e^{2\beta J_{1}+2\beta J_{2}}}{1+e^{2\beta J_{1}+2\beta J_{2}}}\right)^{2}\left(\frac{1-e^{-\beta E_{D}}}{1+e^{-\beta E_{D}}}\right)^{j-i-4}, 
		\label{Eq:correlationvacancyFMFM4}
        \end{align}
    \end{subequations}
	for $i\leq -2$, $j\geq2$. For $i,j\leq -2$ and $i,j\geq2$, $\langle\sigma_{i}\sigma_{j} \rangle$, the correlators  are unchanged by the presence of the vacancy and are given by \eqref{Eq:SpincorrelationFMNN+FMNNN}. All the correlators satisfy symmetry relation $\langle\sigma_{i}\sigma_{j}\rangle = \langle\sigma_{-i}\sigma_{-j}\rangle$. The correlators~\eqref{Eq:correlationvacancyFMFM1}-\eqref{Eq:correlationvacancyFMFM4} are identical to those
	 in the case of antiferromagnetic NNN interactions \eqref{Eq:correlationvacancyFMAFM1}-\eqref{Eq:correlationvacancyFMAFM4}.
	
	The magnetic susceptibility of the vacancy is given by 
	\begin{align}
		\chi(T) - \chi_{0}(T) &= 2\beta\left(\langle\sigma_{-1}\sigma_{1}\rangle - \langle\sigma_{-1}\sigma_{1}\rangle_{0}\right)+4\beta \sum_{j=2}^{\infty}\left(\langle\sigma_{1}\sigma_{j}\rangle-\langle\sigma_{1}\sigma_{j}\rangle_{0}\right) \nonumber\\
		&\quad+4\beta \sum_{j=2}^{\infty}\left(\langle\sigma_{-1}\sigma_{j}\rangle-\langle\sigma_{-1}\sigma_{j}\rangle_{0}\right)+ 2\beta\sum_{i=-2}^{-\infty}\sum_{j=2}^{\infty}\left(\langle\sigma_{i}\sigma_{j}\rangle -\langle\sigma_{i}\sigma_{j}\rangle_{0}\right)
		\nonumber \\
		&\approx 2\beta \left(\frac{1-e^{2\beta J_{2}}}{1+e^{2\beta J_{2}}} - \frac{1-e^{-\beta E_{D}}}{1+e^{-\beta E_{D}}}\right)\nonumber\\
		&\quad + 4\beta \left[\frac{1}{1+e^{2\beta J_{2}}}\,\frac{1-e^{2\beta J_{1}+2\beta J_{2}}}{1+e^{2\beta J_{1}+2\beta J_{2}}}-\frac{1}{1+e^{-\beta E_{D}}}\,\frac{1-e^{-\beta E_{D}}}{1+e^{-\beta E_{D}}}\right]\left(1+e^{\beta E_{D}}\right)\nonumber\\
		&\quad + \frac{\beta}{2}\left[\frac{1-e^{2\beta J_{2}}}{1+e^{2\beta J_{2}}}\left(\frac{1-e^{2\beta J_{1}+2\beta J_{2}}}{1+e^{2\beta J_{1}+2\beta J_{2}}}\right)^{2}-\left(\frac{1-e^{-\beta E_{D}}}{1+e^{-\beta E_{D}}}\right)^{3}\right]\left(1+e^{\beta E_{D}}\right)^{2}
	\end{align}
	At low temperatures, $T \ll 2|J_{1,2}|$,
	\begin{equation}
		\chi(T) - \chi_{0}(T) = -\beta e^{-4\beta J_{1}-6\beta J_{2}}- 2\beta e^{-2\beta J_{1}-6\beta J_{2}} + o\left(\beta e^{-2\beta J_{1}-6\beta J_{2}}\right)
	\end{equation}
 The susceptibility of vacancy is exponentially large. The leading order effect is equivalent to reducing the length of the chain by $b(T)\approx e^{-2\beta J_{1}- 2\beta J_{2}}$.


 \section{Magnetic susceptibilities for finite antiferromagnetic chains}
 \label{Sec:finitechain}
 
  \begin{figure}[hb!]
	\centering
	\includegraphics[width = 3in]{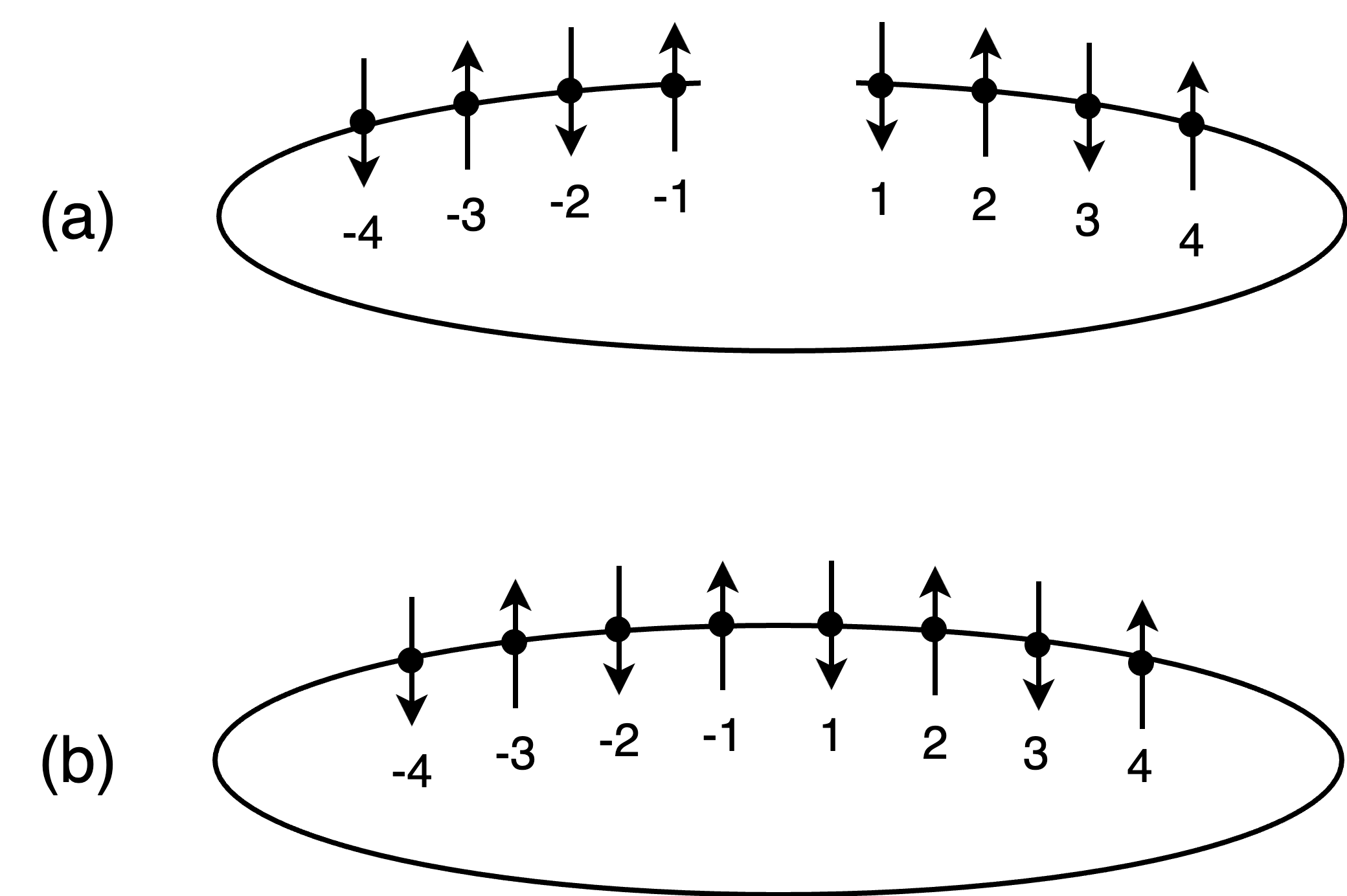}
	\caption{The ground state of Ising chains with antiferromagnetic NN interactions and weaker NNN interactions. (a) A chain with two free ends. (b) A chain with periodic boundary conditions.}
	\label{fig:openchain}
\end{figure}

 This section provides details of the calculation of the magnetic susceptibility of a finite chain with antiferromagnetic NN interactions and weaker NNN interactions ($J_{1}>0$, $|J_{2}|<J_{1}/2$). We assume the length of the chain $N=2n$ to significantly exceed the correlation length $\xi$. In what follows, we label the spins as shown in Fig.~\ref{fig:openchain}a.

The energies of domain walls are modified near the free ends.
The energies of the domain walls between sites $1$ and $2$ and
between sites $-1$ and $-2$ are given by $2J_{1}-2J_{2}$. 

In the limit of a long chain, spin correlators are given by
\begin{subequations}
    \begin{align}
    \langle\sigma_{i}\sigma_{j} \rangle &=0 \quad \text{for}\quad -n\leq i\leq-1,\,\,1\leq j\leq n;\\
    \langle\sigma_{1}\sigma_{j} \rangle &\approx\frac{e^{-2\beta J_{1}+2\beta J_{2}}-1}{1+e^{-2\beta J_{1}+2\beta J_{2}}}\left(\frac{e^{-\beta E_{D}}-1}{1+e^{-\beta E_{D}}}\right)^{j-2} \quad \text{for}\quad 2\leq j\leq n,
\end{align}
\end{subequations}
where $E_{D}=2J_{1}-4J_{2}$ is the energy of a domain wall in the bulk of the chain. For $i,j\leq -1$ and $i,j\geq1$, $\langle\sigma_{i}\sigma_{j} \rangle$, the correlators match those in a chain with periodic boundary conditions (see Fig.~\ref{fig:openchain}b) and are given by Eq.~\eqref{Eq:SpincorrelationAFMNN+FMNNN}. All the correlators satisfy symmetry relation $\langle\sigma_{i}\sigma_{j}\rangle = \langle\sigma_{-i}\sigma_{-j}\rangle$.

The contribution of the free ends to the susceptibility is given by the difference of the susceptibilities of a chain with open boundary conditions and that with periodic boundary conditions:
\begin{align}
    \chi_{\text{open}} - \chi_0 &= 4\beta \sum_{j=2}^{n}\left(\langle\sigma_{1}\sigma_{j}\rangle-\langle\sigma_{1}\sigma_{j}\rangle_{0}\right) + 2\beta\sum_{i=-1}^{-n}\sum_{j=1}^{n}\left(\langle\sigma_{i}\sigma_{j}\rangle -\langle\sigma_{i}\sigma_{j}\rangle_{0}\right)\nonumber\\
    &\approx 2\beta\left(\frac{e^{-2\beta J_{1}+2\beta J_{2}}-1}{1+e^{-2\beta J_{1}+2\beta J_{2}}}-\frac{e^{-\beta E_{D}}-1}{1+e^{-\beta E_{D}}}\right)\left(1+e^{-\beta E_{D}}\right)  -\frac{\beta}{2}\frac{e^{-\beta E_{D}}-1}{1+e^{-\beta E_{D}}}\left(1+e^{-\beta E_{D}}\right)^{2}
\end{align}
At low temperatures, $T \ll 2J_1-2J_{2}, \,2J_1-4J_2$,
\begin{equation}
    \chi_{\text{open}} - \chi_0 = \frac{\beta}{2} + o\left(\beta e^{-2\beta J_{1}+2\beta J_{2}},\beta e^{-\beta E_{D}}\right).
\end{equation}
According to the definition of quasipins in Eq.~\eqref{Eq:QuasispinDefinition}, the leading order contribution corresponds to a quasispin $\sqrt{\left\langle S^{2}\right\rangle}=\frac{1}{2}$ from each of the free ends.


	\twocolumngrid
	

\end{document}